\documentclass[aps,prf,preprint,groupedaddress]{revtex4-2}

\usepackage{graphicx}
\usepackage{natbib}
\usepackage{newtxtext}
\usepackage{newtxmath}
\usepackage{hyperref}
\usepackage{booktabs}
\usepackage{longtable}
\usepackage{lineno}
\hypersetup{
    colorlinks = true,
    urlcolor   = blue,
    citecolor  = black,
}

\newcommand{\RomanNumeralCaps}[1]

\begin{document}

\title{A Unified Numerical Framework for Turbulent Convection and Phase-Change Dynamics in Coupled Fluid-Porous Systems}

\author{Rongfu Guo}
\affiliation{State Key Laboratory for Turbulence and Complex Systems, and Department of Mechanics, School of Mechanics and Engineering Science, Peking University, Beijing 100871, P.R. China}
\author{Yantao Yang}
\email{yantao.yang@pku.edu.cn}
\affiliation{State Key Laboratory for Turbulence and Complex Systems, and Department of Mechanics, School of Mechanics and Engineering Science, Peking University, Beijing 100871, P.R. China}

\date{\today}

\begin{abstract}
This work presents a unified numerical framework for simulating incompressible flows within the coupled fluid-porous-medium system and involving heat and solute transport and phase-changing process. A complete set of governing equations is established based on the Darcy-Brinkman equation, the advection-diffusion equations for heat and solute, and a phase field equation describing the evolution of porous medium. Phase-changing process and relevant influences are incorporated as corresponding source terms. A numerical method is then developed to solve the governing equations. Several different types of model problems are simulated with the numerical method. For the incompressible flows inside a coupled fluid-porous-medium system, the channel turbulence over a porous substrate and the thermal convection in a two-layer system are simulated. For the phase-changing flows, the one-dimensional Stefan problem and the two-dimensional flow of pure water freezing are tested. The results agree with the existing simulations. Finally, the full solver is used to simulate the growth of mushy ice during seawater freezing, which can successfully reproduce the experimental results at the exactly same conditions. Therefore, the developed framework provides a versatile and reliable tool for studying complex multiphase, multi-component transport phenomena in fluid-porous-medium systems involving solid-liquid phase change.
\end{abstract}

\maketitle

\section{Introduction}\label{sec:intro}

Convection and turbulent flows in coupled fluid-porous-medium system are vital processes in many natural environments and engineering applications. Examples include mushy layers in various natural systems~\cite{Worster97,Anderson20}, lava lakes and magmatic systems~\cite{Worster02,Holness17}, metallurgical processes~\cite{Mehrabian70}, drag reduction and heat management~\cite{Hao25,Hung13}. Very often such systems also involve multiple scalar components including heat and concentration fields, and phase-changing processes, such as the mushy layer formed during the sea-ice growth~\cite{Anderson20,Du24}. Therefore, developing a macro scale numerical method for such flows need to model the coupled fluid-porous-medium, the coupled momentum-heat-concentration dynamics, and the phase-changing process. Moreover, the porosity and permeability also evolve depending on the local conditions and strong inhomogeneity appears in the porous region, which represents another challenge in developing numerical methods. 
  
A traditional approach to modeling fluid–porous systems uses separate governing equations: the Navier–Stokes equations in the fluid region and Darcy’s law in the porous medium~\citep{Levy75,Schulze99}. A major challenge in such two-domain methods is defining physically consistent boundary conditions at the interface. Although continuity of normal velocity and pressure is commonly assumed, experiments show these conditions are often inadequate~\citep{Beavers67}. Considerable effort has thus been devoted to developing improved interface models, including stress-jump, velocity-slip, and coupled formulations~\citep{Saffman71,Neale74,Ochoa95,Cieszko99}. Alternatively, a unified single-domain formulation based on the Darcy–Brinkman equations offers a different strategy. Proposed by \citet{Bars06}, this approach incorporates a viscous term in the porous region, effectively creating a smooth transition zone where the Darcy equations remain valid. It ensures automatic continuity of velocity and pressure throughout the domain, eliminating the need for explicit interface conditions. This method performs particularly well when porosity varies smoothly, such as between a mushy layer and melt region, and is also highly suitable for modeling phase-change processes and systems with complex boundary conditions.

A common methodology involves formulating governing equations for the temperature and solute concentration across the entire domain, incorporating a spatially variable porosity field~\citep{Nield17,Hu23}. On this basis, source terms are introduced to represent phase change effects, providing a unified continuum framework that captures both advective and diffusive transport~\citep{Esfahani18,Guo25}. The effective thermal conductivity is typically estimated using mixing models -—- such as series or parallel configurations~\citep{Bhattacharya02}. However, in certain contexts, subgrid-scale enhancement effects become non-negligible, necessitating the introduction of a thermal dispersion correction to account for additional heat transfer caused by tortuous flow paths and velocity fluctuations within the porous medium~\citep{Pedras08,Yang10}. For solute transport in mushy layers, the effective diffusivity is often approximated as being proportional to the local porosity~\citep{Feltham06,Pedras08}. However, more refined models must also account for the tortuous microstructure of the solid matrix, which elongates the diffusion path and reduces transport efficiency~\citep{Lange10,Chen14}. This is commonly described using an Archie’s law-type formulation~\citep{Niu18}, where the effective diffusivity scales with porosity raised to an exponent greater than one, or through the explicit incorporation of a tortuosity factor that reflects the local pore geometry~\citep{Wu19}.

In order to model the phase change within mushy layers, the entire mushy zone must satisfy the liquidus relation for a binary mixture throughout the entire domain, ensuring consistency with local thermodynamic equilibrium~\citep{Kerr90a,Feltham06}. The presence of supercooling further influences interface kinetics and morphology, and must be incorporated to accurately capture non-equilibrium phase transition behavior~\citep{Kerr90b,Wettlaufer97}. Phase-field methods provide a diffuse-interface representation for tracking phase distribution and evolution. Classical formulations employ the Cahn–Hilliard equation~\citep{Ding07}, which contains a fourth-order term representing surface tension. This term imposes severe numerical constraints, including extremely small time steps and high computational cost, and can lead to nonphysical interface evolution under certain conditions~\citep{Yue07}. Alternatively, a second-order phase-field formulation has been developed~\citep{Chen25}, which circumvents these issues by avoiding high-order derivatives, thereby permitting larger time steps and eliminating spurious numerical artifacts while preserving the essential physics of phase evolution.
  
In this study, we develop a unified numerical framework for simulating turbulent thermal convection and phase-change dynamics in coupled fluid-porous systems, which can effectively handle systems with different solid and liquid thermal conductivities and accommodates spatiotemporally varying porosity. The method is built based on the Darcy-Brinkman formulation with a modified phase-field method, achieving smooth two-way coupling across evolving interfaces. A unified set of governing equations is established for momentum, energy, solute transport, and phase evolution within a single-domain continuum description. The model is then rigorously validated against a series of benchmark cases.

The structure of this paper is organized as follows. $\S$\ref{sec:mod} provides the details about the unified governing equations for the coupled fluid-porous system. The numerical methodology, including the temporal discretization scheme, the operator-splitting algorithm for the momentum equation, and the projection method, are described in $\S$\ref{sec:method}. Validations of the model against a series of benchmark problems are presented in $\S$\ref{sec:cases}. Finally, $\S$\ref{sec:con} summarizes the main conclusions and discusses potential applications of the developed framework.

\section{Mathematical formulation and governing equations}\label{sec:mod}

\subsection{The flow configuration}

\begin{figure}
  \centering
  \includegraphics[width=0.7\textwidth]{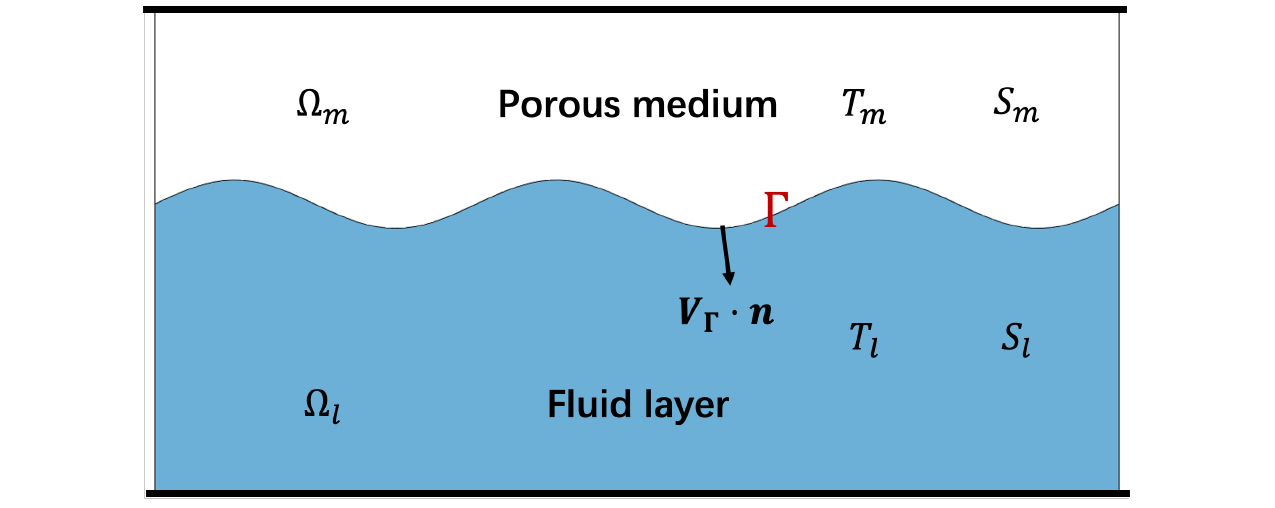}
  \caption{Schematic illustration of the solidification problem involving a binary fluid. The mushy domain $\Omega_m$ is separated from the liquid domain $\Omega_l$ by the liquid-solid interface $\Gamma$. The temperature $T_{l,m}$ and the concentration $S_{l,m}$ are solved in the two domains.}\label{fig:sketch}
\end{figure}

The flow domain considered in the present study consists of a layer of pure fluid and a layer of porous medium, as shown in figure~\ref{fig:sketch}. When gravity is considered, the two layers are usually stacked along the direction of gravity, which is in the opposite direction of the $z$-axis. The two horizontal directions are denoted by $x$ and $y$, respectively. However, it should be pointed out that the orientation of the two layers is not limited to the situation shown in figure~\ref{fig:sketch}, but depends on the actual dynamics or is prescribe in advance. Denote the porous mushy region by $\Omega_m$ and the liquid region by $\Omega_l$, respectively. The whole flow domain is then $\Omega=\Omega_l\cup\Omega_m$. A phase variable $\phi$ is introduced to represent local porosity or the fraction of fluid. $\phi$ varies smoothly from $0$ to $1$ corresponds to the transition from pure solid to pure fluid. A value of $\phi$ between $0$ and $1$ represents the local porosity.

To model this coupled system, we adopt a unified single-domain approach based on the Darcy-Brinkman formulation for momentum transport. This framework automatically ensures the continuity of velocity and pressure across the entire domain, effectively bypassing the need for explicit interface conditions that often pose challenges in two-domain methods. Beside the momentum, energy, and concentration equations, a modified phase-field method is integrated to describe the evolution of $\phi$. So that it is possible to handle the high solid-to-fluid thermal conductivity contrast, and the spatiotemporal variation of porosity within the computational domain. 

In the following subsections we will respectively describe the phase-field equation, the momentum equation, the energy equation, and the concentration equation. Then the whole set of non-dimensional governing equations and the resulted control parameters will be summarized.

\subsection{The modified phase-field equation}\label{subsec:pe_evo}

As mentioned before, a phase field $\phi$ is introduced to track the evolution of liquid and porous regions. In the porous layer, it also describes the local variation in porosity. For stationary or known porous layer, the porosity can be preset without the need of solving the equation. For dynamic mushy layers such as those formed during seawater freezing, the following advection-diffusion equation is employed
\begin{equation}\label{eq:phi_dim}
  \partial_t\phi + \mathbf{U}\cdot\nabla\phi = \kappa_\phi\nabla^2\phi +\mathcal{G}\left(T-T_\phi\right).
\end{equation}
Here $\mathbf{U}$ is the Darcy velocity inside the porous layer or the fluid velocity in the liquid layer. $\kappa_\phi$ is the diffusion coefficient which is set at a sufficiently small positive value. The last term in the equation represents the variation of porosity caused by the phase-changing process inside the porous layer. $T_\phi=T_0-\lambda S$ is the local liquidus temperature, or the freezing temperature at solutal concentration $S$. $T_0$ is the freezing temperature of pure liquid. $\lambda$ is the equilibrium temperature offset coefficient or the local slope of the liquidus line. The parameter $\mathcal{G}$ controls the response rate of phase transition kinetics. 

Some extra comments should be made. Unlike the normal phase-field equation such as the Cahn-Hilliard equation~\cite{Liu21}, the phase-field equation used here does not include a double-well potential term and the fourth-order anti-diffusion term. The former induces the phase separation inside the mixture while the latter keeps the interfaces sharp. However, since we are interested in the phase-changing process in the mushy porous layer, neither effects are needed in the current formulation.

\subsection{The unified momentum equation}

In the current flow system, the porosity of phase field $\phi$ can vary significantly from the value $\phi=1$ for pure liquid to $\phi=0$ for pure solid. Therefore, the inertial and viscous effects must be included in the momentum equation, and we employ the Darcy-Brinkman equation derived from the volume-averaging method~\cite{Bars06}, which reads
\begin{equation}
  \partial_{t}\mathbf{U} = -\mathbf{U}\cdot\nabla\left(\frac{\mathbf{U}}{\phi}\right) 
      - \frac{1}{\rho_{\rm ref}}\left[\nabla P_l\right]
      + \nu\left(\nabla^{2}\mathbf{U} -\frac{\phi}{\mathcal{K}}\mathbf{U}\right) 
      - \frac{g\phi\rho'}{\rho_{\rm ref}}\mathbf{e}_z.
\end{equation}
Here, $P_l$ is pressure in the liquid phase, $\rho_{\rm ref}$ is the liquid density at the reference state. The square bracket $[\cdot]$ stands for the volume averaging within the representative element volume (REV) of porous medium. Therefore, this term denotes the macroscopic pressure gradient which drives the macroscopic velocity $\mathbf{U}$. The Boussinesq approximation has been used so that the buoyancy force is proportional to the density anomaly $\rho'$ and of course the porosity $\phi$, with $g$ being the gravitational acceleration. In the Darcy resistance term, the permeability $\mathcal{K}(\phi)=K_0 f(\phi)$ is assumed to be a scalar function of porosity $\phi$, and its explicit form depends on the specific problem. The current study also is confined to the incompressible flows, saying
\begin{equation}\label{eq:continu_dim}
  \nabla\cdot\mathbf{U} = 0.
\end{equation}

Following the similar strategy as in~\citet{Bars06}, here we provide a more simplified pressure-gradient term. According to volume-averaging theory, $[\nabla P_l]$ can be expanded as:
\begin{equation}\label{eq:pressure}
  \left[\nabla P_l\right] = \nabla\left[P_l\right] + \frac{1}{\Delta V}\int_{\Delta A} P_l\mathbf{n}_l \,dA = \nabla\left[P_l\right] + \frac{\left[P_l\right]}{\Delta V}\int_{\Delta A} \mathbf{n}_l \,dA + \frac{1}{\Delta V}\int_{\Delta A} \widehat{P_l} \mathbf{n}_l \,dA.
\end{equation}
The last term, which represents the pressure fluctuations on the solid-liquid interface with respect to the volume-averaged pressure, is assumed to be small and negligible as done in~\citet{Bars06}. Moreover, by assuming that the solid–liquid distribution within the averaging volume $\Delta V$ is locally isotropic, the surface integral of the normal vector $\mathbf{n}_l$ should also be negligible, namely,
\begin{equation}
  \int_{\Delta A} \mathbf{n}_l \,dA \approx 0.
\end{equation}
Then the second to last term in~\eqref{eq:pressure} is small and one has
\begin{equation}
  \left[\nabla P_l\right] \approx \nabla\left[P_l\right] = \nabla\left(\phi\left[P_l\right]^l\right),
\end{equation}
with $[\cdot]^l$ being the average over the liquid phase within REV. It should be pointed out that in~\citet{Bars06} the pressure term is simplified as $[\nabla P_l] = \phi \nabla[p_l]^l$. Here we show that the difference between the two forms is negligible.

Define the macroscopic pressure as $P=\phi{\left[P_l\right]}^l$, then the final form of momentum equation reads
\begin{equation}\label{eq:mom_dim}
  \partial_{t}\mathbf{U} = -\mathbf{U}\cdot\nabla\left(\frac{\mathbf{U}}{\phi}\right) 
      - \frac{1}{\rho_{\rm ref}}\nabla P
      + \nu\left(\nabla^{2}\mathbf{U} -\frac{\phi}{\mathcal{K}}\mathbf{U}\right) 
      - \frac{g\phi\rho'}{\rho_{\rm ref}}\mathbf{e}_z.
\end{equation}
Here, the term $\nabla P$ represents the macroscopic effect of the pressure gradient when considering the volume fraction $\phi$ occupied by liquid phase. It is readily to show that the Darcy-Brinkman equation~\eqref{eq:mom_dim} recovers the standard Navier-Stokes equation in the pure liquid region with $\phi=1$ and $K\to\infty$. On the other hand, in the dense porous region the viscous stresses are much smaller than the Darcy resistance, and the inertial terms are negligible, the equation asymptotically approaches the Darcy equation
\begin{equation}\label{eq:darcy_dim}
  \mathbf{U} = -\frac{\mathcal{K}}{\nu}\left(\frac{\nabla P}{\rho_{\rm ref}} + g\mathbf{e}_z\right),
\end{equation}
where $P$ representing the microscopic fluid pressure in accordance with conventional Darcy's law, which differs from the macroscopic pressure $P$ in equation~\eqref{eq:mom_dim}. Therefore, equation~\eqref{eq:mom_dim} provides a unified description for the incompressible flow with the flow domain consisting of pure fluid and porous medium.

\subsection{The temperature equation}\label{subsec:eng_eqn}

In the coupled fluid-porous-medium systems, both solid and liquid phases participate in the heat transfer. Here we assume the mixture is in the local thermal equilibrium, i.e., the solid and liquid materials have the same local temperature $T$. Denote the volumetric heat capacities of solid and liquid phases by $c_s$ and $c_l$, and the thermal conductivities by $k_s$ and $k_l$, respectively. Then the temperature equation read
\begin{equation}\label{eq:temp_dim}
  \partial_{t}\left(c_m T\right) = -c_f\mathbf{U}\cdot\nabla T 
       + \nabla\cdot\left(k_e\nabla T\right) -\mathcal{L}\partial_t\phi,
\end{equation}
with $c_m=\phi c_l + (1-\phi) c_s$. The effective thermal conductivity $k_e$ combines the static part $k_m=\phi k_l + (1-\phi) k_s$ which is the volume averaging of solid and liquid phases, and an extra term $k_{dis}$ which represents the enhancement of thermal dispersion due to the subgrid-scale flows at the high porosity regions inside the porous layer. The latter is assumed to be linearly proportional to the magnitude of velocity $\mathbf{U}$. The effective thermal conductivity is then calculated as
\begin{equation}\label{eq:keff}
  k_e=k_m+k_{dis} = \phi k_l + (1-\phi) k_s + \epsilon_{dis}(\phi) c_m \left|\mathbf{U}\right|.
\end{equation}
The coefficient $\epsilon_{dis}$ depends on the local porosity $\phi$ and ensures a smooth transition from the porous medium region to liquid region. Generally, $\epsilon_{dis}$ takes a positive constant for $\phi\le0.9$ and rapidly decreases to zero as $\phi$ exceeds $0.9$ and increases to unity. The specific function form of $\epsilon_{dis}$ will be given in the corresponding problems in the next section.

The last term in equation~\eqref{eq:temp_dim} represents the latent heat due to the phase-changing. $\mathcal{L}$ is the latent heat per unit volume. While $\partial_t \phi$ is exactly the temporal increasing rate for the volume fraction of liquid, therefore the negative sign of the term. In our numerical framework, we use the following equivalent equation for temperature $T$ as
\begin{equation}\label{eq:temp_dim_final}
  \partial_{t}T =-\frac{c_f}{c_m}\mathbf{U}\cdot\nabla T 
    + \frac{1}{c_m}\nabla\cdot\left(k_e \nabla T\right) -\frac{\mathcal{L}}{c_m}\partial_t\phi -\frac{T}{c_m}\partial_{t}c_m.
\end{equation}

\subsection{The solute equation}\label{eq:solute_dim}

Solute transport in porous media occurs exclusively through the fluid phase, and the transport equation can be written as
\begin{equation}\label{eq:solu_dim}
  \partial_{t}\left(S\phi\right) = -\mathbf{U}\cdot\nabla S+\nabla\cdot\left(\kappa_S^e\nabla S\right). 
\end{equation}
Here $S$ is the solute concentration in liquid phase. The effective solute diffusivity $\kappa_S^e$ incorporates the influence of porous micro-structures on local solute diffusion. Specifically, we set
\begin{equation}
  \kappa_S^e = \kappa_S \frac{\phi}{\tau},  \quad \mbox{with~}
  \tau=\phi^{-\gamma},
\end{equation}
in which$\kappa_S$ is the molecular diffusivity of solute in the liquid phase. $\tau$ is the tortuosity of micro-structure and modeled by using Archie's law with the exponent $\gamma$ reflecting the geometric complexity of microstructure~\cite{Chen14}. When $\gamma=0$, the tortuosity effect is absent and the effective diffusivity varies linearly with porosity, which is usually the case for homogeneous porous medium. However, in porous medium such as mushy ice, the highly complex microstructure and significantly tortuous flow paths lead to a suppression effect of solute diffusion. In this study, we adopt the value $\gamma=1$, which corresponds to a second-order power-law relation between $\kappa_S^e$ and $\phi$.

For the convenience of numerical method, we also cast equation~\eqref{eq:solu_dim} to the following form
\begin{equation}\label{eq:solu_dim_final}
  \partial_{t}S =-\frac{1}{\phi}\mathbf{U}\cdot\nabla S +\frac{1}{\phi}\nabla\cdot\left(\kappa_S^e\nabla S\right) -\frac{S}{\phi}\partial_t\phi.  
\end{equation}
And it is evident that when $\phi=1$ in the liquid layer, the above equation recovers the standard advection-diffusion equation for solute concentration.

\subsection{The complete set of nondimensionalized governing equations}\label{subsec:goveqn-ndim}

We now give the complete set of governing equations, then introduce the nondimensionalization and the resulting nondimensional control parameters. The governing equations include the phase-field equation~\eqref{eq:phi_dim}, the continuity equation of incompressible flow~\eqref{eq:continu_dim}, the momentum equation~\eqref{eq:mom_dim}, the temperature equation~\eqref{eq:temp_dim_final}, and the solute equation~\eqref{eq:solu_dim_final}. When the buoyancy force is considered, an equation of state must be used to relate the density anomaly to temperature and solute concentration, namely $\rho'=\rho'(T,~S)$. A common choice for liquid is the linear equation of state as $\rho'=\rho_{\rm ref}(-\beta_T T + \beta_S S)$, with $\beta_T$ and $\beta_S$ being the respective thermal expansion or solutal contraction coefficients. However, when the phase-changing process is included for water, the density inversion effect or a nonlinear equation of state should be used. In this section, we use the linear equation of state as an example, while in Section~\ref{sec:cases} when the freezing and melting of water-ice is considered, the corresponding nonlinear equation of state will be given there.

Let $H$ be the characteristic scale for length, $U$ for velocity, $\Delta_T$ for temperature, and $\Delta_S$ for solute concentration, respectively. The common choices for buoyancy-driven flows are the total layer height, the free-fall velocity, and the temperature and solute differences across the layer. All the flow quantities can be nondimensionalized accordingly. Especially, the nondimensional temperature and solute anomalies are $\theta=(T-T_{\rm ref})/\Delta_T$ and $s=(S-S_{\rm ref})/\Delta_S$. For buoyancy-driven flows, $H$ is usually the total height of domain, $U=\sqrt{g\beta_T\Delta_TH}$ the free-fall velocity set by the temperature field, and $\Delta_T$ and $\Delta_S$ the scalar differences across the domain height, respectively. Then the nondimensional governing equations read
\begin{subequations}\label{eq:goveqn}
\begin{align}
    & \nabla\cdot\mathbf{u} = 0, \label{eq:con-ndim}	\\
    & \partial_{t}\mathbf{u} = -\mathbf{u}\cdot\nabla\left(\frac{\mathbf{u}}{\phi}\right) 
         - \nabla p 
         + \sqrt{\frac{Pr}{Ra}}\left(\nabla^{2}\mathbf{u} - \frac{\phi}{Da\,f} \mathbf{u} \right) 
         + \phi\left(\theta -\Lambda s\right)\mathbf{e}_z, \label{eq:db-ndim} \\
    & \partial_{t}{\theta} = -\frac{\mathbf{u}\cdot\nabla{\theta}}{\alpha_c} 
         + \frac{\nabla\cdot\left(\alpha_k \nabla\theta\right)}{\alpha_c\sqrt{Pr Ra}} 
         - \frac{St}{\alpha_c}\partial_{t}\phi 
         - \frac{\theta}{\alpha_c}\partial_t\alpha_c, \label{eq:te-ndim}	\\
    & \partial_{t}s = -\frac{\mathbf{u}\cdot\nabla s}{\phi} 
         + \frac{\nabla\cdot\left(\alpha_s \nabla s\right)}{\phi Le\sqrt{Pr Ra}} 
         - \frac{s}{\phi}\partial_{t}\phi, \label{eq:sa-ndim} \\ 
    & \partial_t\phi = -\mathbf{u}\cdot\nabla\phi + C_\phi\nabla^2\phi 
         + C_G \left(\theta + C_\lambda s - \theta_0\right). \label{eq:pe-ndim}
\end{align}
\end{subequations}
Note that $\alpha_k=k_e/k_l$, $\alpha_c=c_m/c_l$, and $\alpha_s=\phi^2$ vary with time and spatial coordinates and should be updated during simulation. The two nondimensional coefficients $C_\phi = \kappa_\phi/H$ and $C_G=\mathcal{G}H\Delta_T/U$ need to be adjusted to properly reflect the corresponding physical processes. And other nondimensional parameters include
\begin{equation}
  Pr = \frac{\nu}{\kappa_T},  \quad
  Le = \frac{\kappa_T}{\kappa_S},  \quad
  Ra = \frac{g\beta_T\Delta_T H^3}{\kappa_T\nu},  \quad
  Da = \frac{K_0}{H^2},  \quad
  \Lambda = \frac{\beta_S\Delta_S}{\beta_T\Delta_T},  \quad
  St = \frac{\mathcal{L}}{c_p\Delta_T},  \quad
  C_\lambda = \frac{\lambda \Delta_S}{\Delta_T}.
  \nonumber
\end{equation}

The above set of governing equations, together with proper boundary conditions, will be used to simulate the dynamics of buoyancy-driven flows or the turbulent flows within fluid-porous-medium system, with or without phase-changing process. In the following two sections we first describe the numerical schemes and then apply the framework to several representative model problems.

\section{The numerical methods}\label{sec:method}

In this section we present the numerical method which is used to solve the governing equations~\eqref{eq:goveqn}. The method is built upon the well-developed in-house solver which has been widely applied to wall-bounded turbulence and convection flows~\cite{Verzicco96,Zhu18}. We also incorporate the multi-grid technique~\cite{Ostilla15} to efficiently resolve the concentration field which usually has very small molecular diffusivity and requires very fine resolution. Note that we solve the phase field and concentration field on the refined mesh, and velocity and temperature on the base mesh, respectively. Therefore, interpolation is needed for the information exchange between the two sets of meshes.

\subsection{The overall procedure}

We first give the overall procedure for updating the phase field $\phi$, velocity $\mathbf{u}$, temperature $\theta$, and concentration $s$ during each time step. The key steps are listed as follows.
\begin{itemize}
  \item[(1)] \textbf{Phase Field Update:} Solve the phase field equation\eqref{eq:pe-ndim} on the refined grid to update the porosity from $\phi^n$ to $\phi^{n+1}$, determining the distribution of liquid and mushy zones for the current time step. The results are then interpolated to the base grid.
  \item[(2)] \textbf{Update location-dependent properties:}
  \begin{itemize}
    \item[(2.1)] Compute the dimensionless effective thermal conductivity $\alpha_k$ cell-wise on the base grid using equation\eqref{eq:te-ndim}, based on $\phi^{n+1}$ and $\mathbf{u}$.
    \item[(2.2)] Compute the dimensionless effective solute diffusivity $\alpha_s$ cell-wise on the refined grid using equation\eqref{eq:sa-ndim}, based on $\phi^{n+1}$.
    \item[(2.3)] Calculate the Darcy number $Da$ for each cell using the Kozeny–Carman relation, and subsequently determine the Darcy resistance term $\eta$.
    \item[(2.4)] Compute the seepage velocity in the mushy zone as $\mathbf{u}_l^n = \mathbf{u}^n / \phi^{n+1}$.
  \end{itemize}
  \item[(3)] \textbf{Velocity Update:}
  \begin{itemize}
    \item[(3.1)] \textbf{Velocity-Pressure Coupling:} Solve the unified Darcy–Brinkman momentum equation\eqref{eq:db-ndim} on the base grid by using the updated physical properties. The standard fractional time-step method is used to obtain velocity $\mathbf{u}^{n+1}$ and pressure $p^{n+1}$. 
    \item[(3.2)] \textbf{Velocity Field Prolongation:} Interpolate the newly updated velocity field $\mathbf{u}^{n+1}$ from the base grid to the refined grid to support subsequent solute calculations on the refined grid.
  \end{itemize}
  \item[(4)] \textbf{Scalars Update:}
  \begin{itemize}
    \item[(4.1)] \textbf{Temperature:} Solve the temperature transport equation\eqref{eq:te-ndim} on the base grid to update the temperature field to $\theta^{n+1}$, then interpolate the temperature field to the refined grid for use in the phase field update.
    \item[(4.2)] \textbf{Solute:} Solve the solute transport equation\eqref{eq:sa-ndim} on the refined grid to advance the solute field to $s^{n+1}$, then interpolate the solute field back to the base grid for use in the velocity field solution.
  \end{itemize}
  \item[(5)] \textbf{Boundary Condition Update:} Update the boundary conditions for the velocity, temperature, and solute fields based on the current time step results.
\end{itemize}

\subsection{The discretization schemes}\label{subsec:dis_spa}

The discretization scheme is very similar to that in~\citet{Ostilla15} and based on the staggered grids. The fractional time-step method is used and the temporal integration utilizes a third-order Runge-Kutta method. For the advection terms we use the second-order upwind scheme, while other terms is discretized by the second-order central difference scheme. The nonlinear terms and source terms are treated explicitly by an Adams-Bashforth type of scheme and the diffusion terms semi-implicitly by an Crank-Nicholson type of scheme, respectively. The divergence-free condition of velocity is enforced by the projection step which requires solving a Poisson equation. In our setup, the periodic condition is usually adopted in our flow systems, the Fast-Fourier-Transform can be employed in the horizontal directions in the Poisson solver. The resulted linear systems due to the semi-implicit treatment of diffusion terms are solved by the factorization method.

Special treatments are needed for the extra terms introduced in the governing equations~\eqref{eq:goveqn}. For most of the property parameters such as viscosity and diffusivity, one needs to calculate values at cell faces from cell centers, and the harmonic mean is used to increase the numerical stability since these quantities may have very sharp variation in space. The linear Darcy resistance term in~\eqref{eq:db-ndim} is treated implicitly. All the cross source terms are treated explicitly. Note that temporal derivatives are involved in several source terms. These terms are calculated by the first-order forward differencing.

Specifically, the last source term in the phase-field equation~\eqref{eq:pe-ndim} is discretized as
\[
(Q_\phi)^n = C_G (\theta^n + C_\lambda s^n - \theta_0).
\]
For the temperature equation~\eqref{eq:te-ndim} and the solute equation~\eqref{eq:sa-ndim}, both the effective diffusivities $\alpha_k$ and $\alpha_s$ vary spatially and temporally. In order to efficiently treat the spatiotemporal variation of diffusivities, the temperature and solute equations are cast into the following forms
\begin{equation}\label{eq:ts_num}
  \partial_t\theta = -\frac{\nabla\cdot\mathbf{\xi}_T}{\alpha_c} +\frac{\alpha_\kappa^{\max} \nabla^2 \theta}{\sqrt{Pr Ra}} - Q_T, \qquad
  \partial_t s = -\frac{\nabla\cdot\mathbf{\xi}_S}{\phi} +\frac{\nabla^2 s}{Le \sqrt{Pr Ra}} - Q_S,
\end{equation}
with 
\begin{equation*}
  \mathbf{\xi}_T = \mathbf{u}\theta - \frac{\alpha_k -\alpha_c\alpha_\kappa^{\max}}{\sqrt{Pr Ra}}\nabla\theta, \qquad
  \mathbf{\xi}_S = \mathbf{u} s - \frac{\alpha_s - \phi}{Le \sqrt{Pr Ra}} \nabla s.
\end{equation*}
Then the two terms $\mathbf{\xi}_T$ and $\mathbf{\xi}_S$ are treated explicitly as advection terms. The diffusion terms in~\eqref{eq:ts_num} have spatially constant diffusivities and the normal semi-implicit method can be used. The source terms are discretized as
\begin{equation*}
  {\left(Q_T\right)}_i^n =\frac{St}{1-c_i/c_f}\frac{\ln\left({\alpha_c}_i^{n+1}/{\alpha_c}_i^n\right)}{\Delta t} +\theta_i^n\frac{\ln\left({\alpha_c}_i^{n+1}/{\alpha_c}_i^n\right)}{\Delta t},
  \qquad
  {\left(Q_S\right)}_i^n =s_i^n\frac{\ln\left({\phi_i^{n+1}/\phi_i^n}\right)}{\Delta t}.
\end{equation*}
  
\section{Validations of the numerical method}\label{sec:cases}

We now test the numerical method developed in the previous section. Different types of model flows are simulated and compared with existing simulations and experiments to validate the different aspects of method. The Darcy-Brinkman flow solver will be tested by the canonical channel flow over a porous region in subsection~\ref{subsec:channel}. The coupling between momentum and temperature fields is then validated by simulating the convection flow inside the fluid-porous-medium two-layer system in subsection~\ref{subsec:conv}. In subsection~\ref{subsec:phase} the phase-changing simulations are carried out for the one-dimensional (1D) Stefan problem and the two-dimensional (2D) freezing of pure water. And finally, the full solver is tested by simulating the process of seawater freezing and the development of mushy ice layer in subsection~\ref{subsec:mushyice}. For reader's convenience, the exact form of the governing equations for the specific problem are given in each subsection.

\subsection{The channel flow over a permeable substrate}\label{subsec:channel}

\begin{figure}
  \centering
  \includegraphics[width=0.6\textwidth]{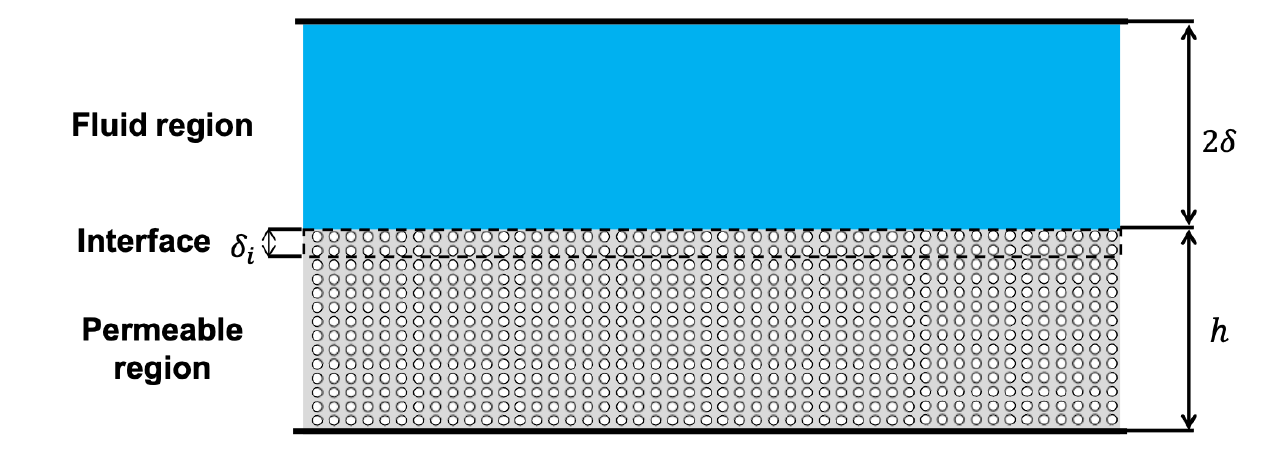}
  \caption{Schematic of the turbulent channel flow over a permeable substrate following~\citet{Breugem06}. The channel has a height of $2\delta$, bounded by an impermeable wall at the top and a porous layer of thickness $h=2\delta$ and porosity $\phi$ at the bottom. An interfacial layer of thickness $\delta_i$ facilitates a smooth transition in porosity and permeability.}\label{fig:chan_sketch}
\end{figure}

We first use the developed method to simulate the turbulent channel flow over a porous boundary. The domain configuration is the same as that in~\citet{Breugem06} and shown in figure~\ref{fig:chan_sketch}. The channel has the height of $2\delta$ and lies above a porous medium of the height of $h=2\delta$. The porosity $\phi$ is uniform inside the porous layer, with a transition layer of thickness $\delta_i$ at the top of porous layer within which the porosity transits smoothly from the fluid layer to the homogeneous porous layer. The governing equations for this flow read
\begin{subequations}
  \begin{align}
    &\nabla\cdot\mathbf{u} =0,   \\
    &\partial_{t}\mathbf{u}+\mathbf{u}\cdot\nabla\left(\frac{\mathbf{u}}{\phi}\right) 
         = -\nabla p+\nu\left(\nabla^{2}\mathbf{u}-\frac{\phi}{Da\,f}\mathbf{u}\right) + \mathbf{f}_{b},
  \end{align}
\end{subequations}
where the dimensionless permeability function follows the Kozeny-Carman relation
\begin{equation}
  f(\phi) = \frac{\phi^3}{{\left(1-\phi\right)}^2}, \qquad 
  Da = \frac{K_0}{H^2}\quad \mathrm{with}\quad K_0=\frac{D^2}{180}.
\end{equation}
The flow rate is maintained at constant by adjusting the body force $\mathbf{f}_{b}$.

\begin{figure}
  \centering
  \includegraphics[width=\textwidth]{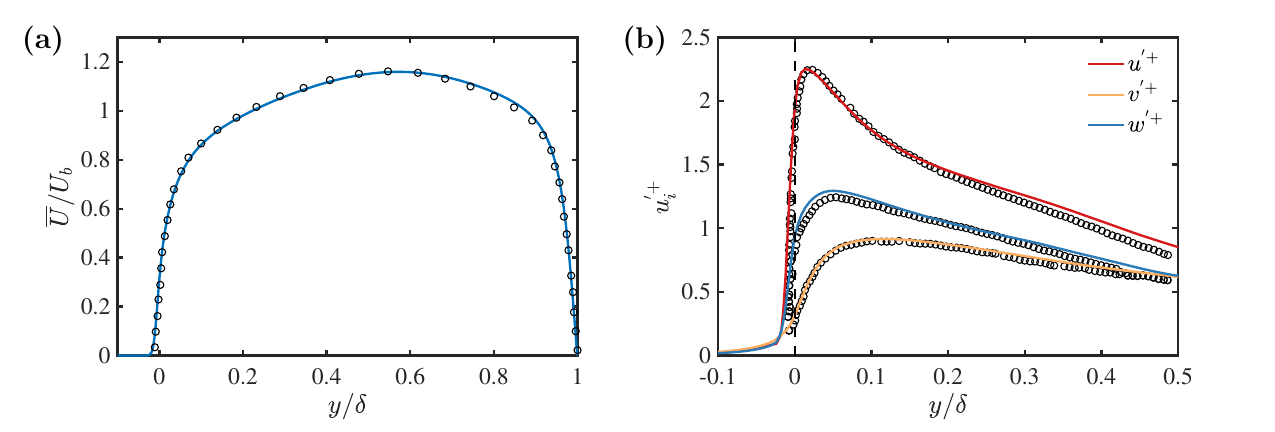}
  \caption{Validation of the mean and fluctuating velocity profiles for channel flow over a permeable substrate with porosity $\phi=0.8$.\,(a) Mean velocity profile normalized by the bulk velocity $U_b$ as a function of dimensionless height $z/H$.\,(b) Root mean square velocity fluctuations in wall units. Symbols denote the reference data from \citet{Breugem06} obtained using the Volume-Averaged Navier-Stokes (VANS) equations.}\label{fig:chan_stats}
\end{figure}

We simulate the case E80 of~\citet{Breugem06} with $\phi=0.8$ and $D=10^{-3}$. The bulk Reynolds number is fixed at $Re_b=5500$. The domain size is $L_x \times L_y \times L_z = 5 \times 3 \times 2$. The corresponding grid size is $N_x\times N_y \times N_z = 256 \times 256 \times 192$. Figure~\ref{fig:chan_stats} compares the mean profiles of streamwise velocity and root-mean-square (rms) of three fluctuation velocity components between our results and those reported in~\citet{Breugem06}. The agreement is very well.
  
\subsection{Convection in a fluid-porous-medium two-layer system}\label{subsec:conv}

\begin{figure}
  \centering
  \includegraphics[width=0.6\textwidth]{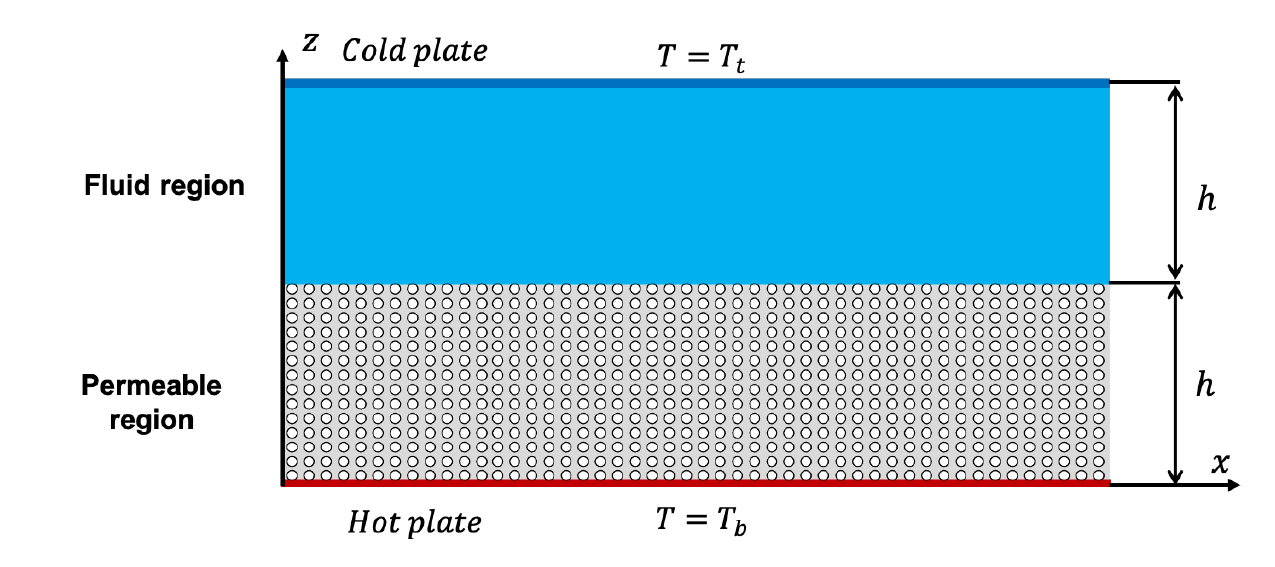}%
  \caption{Schematic of the two-layer convection system. The domain consists of a fluid-saturated porous medium of depth $h$ in the lower half ($-h\le z<0$) and a free fluid layer of depth $h$ in the upper half ($0<z\le h$). The system is driven by a temperature difference between the hot bottom ($T_b$) and cold top ($T_t$) boundaries.}\label{fig:conv_sketch}
\end{figure}

The second model flow is the buoyancy-driven convection flow in a fluid-porous-medium two-layer system as shown in figure~\ref{fig:conv_sketch}. Both the fluid and porous layers have the height of $h$ and the two layers are stacked in the vertical direction. The whole domain is heat from below and cooled from above. The temperature at the top and bottom plates is fixed at $T_t=0$ and $T_b=1$, respectively. The configuration is the same as that used in~\citet{Reun21}. The governing equations read
\begin{subequations}
\begin{align}
    &\nabla\cdot\mathbf{u} = 0,	\\
    &\partial_{t}\mathbf{u}+\mathbf{u}\cdot\nabla\left(\frac{\mathbf{u}}{\phi}\right) 
    = -\nabla p +\sqrt{\frac{Pr}{Ra}}\left[\nabla^{2}\mathbf{u}-\frac{\phi}{Da\,f}\mathbf{u}\right] +\phi\theta\mathbf{e}_z,  \\
    &\partial_{t}{\theta} +\mathbf{u}\cdot\nabla{\theta} =\frac{\nabla^2\theta}{\sqrt{Pr Ra}},
\end{align}
\end{subequations}

\begin{figure}
  \centering
  \includegraphics[width=0.9\textwidth]{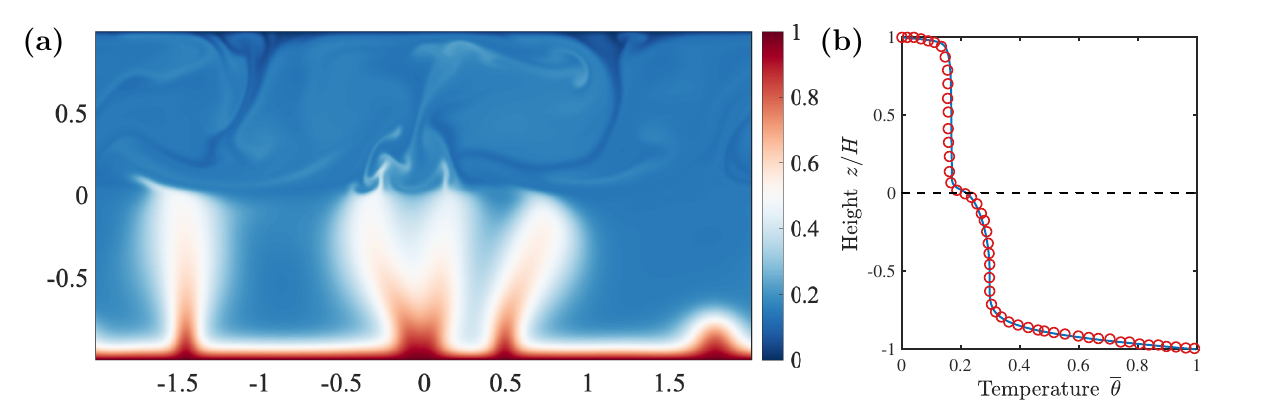}%
  \caption{Numerical results for coupled convection in a fluid-porous layer. (a) Snapshot of the temperature field $\theta$.\,(b) Comparison of the mean temperature profile between our results and those from~\citet{Reun21}.}\label{fig:conv_field}
\end{figure}

\begin{figure}
  \centering
  \includegraphics[width=\textwidth]{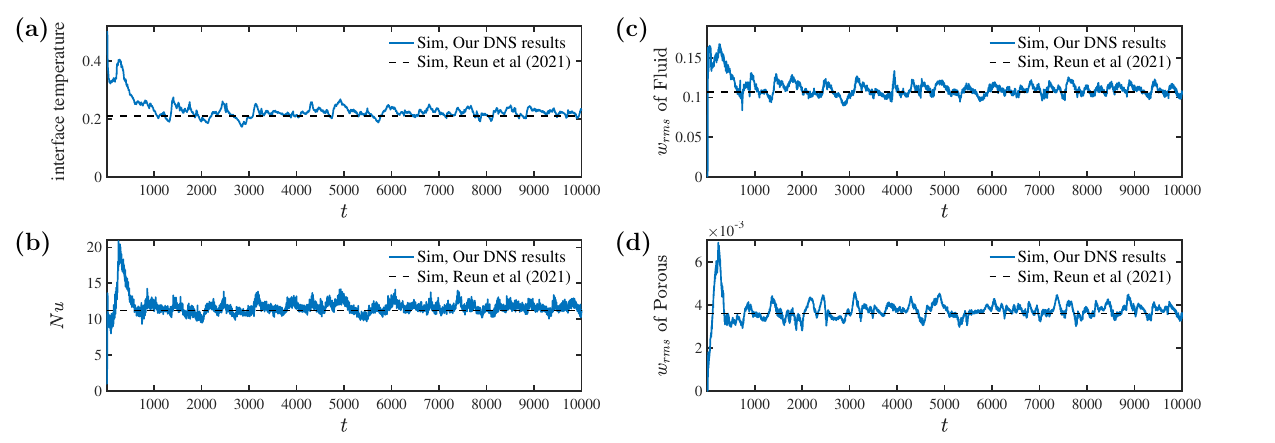}%
  \caption{Temporal evolution of (a) interface temperature, (b) global Nusselt number, (c) rms vertical velocity in the fluid layer, and (d) rms vertical velocity in the porous layer. The horizontal dashed lines mark the values reported in~\citet{Reun21}.}\label{fig:conv_stats}
\end{figure}

The parameters are chosen as $Pr=1$, $Ra=10^8$, $Da=10^{-5.5}$. The porosity is set at $\phi=1$. We set function $f^{-1}(z)$ equal to unit in the porous region and zero in the fluid region. The domain width is $L_x=4h$ with the resolution of $N_x\times N_z = 512\times 384$. And the results are compared to the case with exactly same parameters in~\citet{Reun21}. Figure~\ref{fig:conv_field}(a) plots the typical flow field of the convection in two-layer flows. The different scales of thermal plumes in the lower porous layer and the top fluid layer are clearly visible. Figure~\ref{fig:conv_field}(b) compares the mean temperature profile of our simulation with that given in~\citet{Reun21}. The agreement is very well. The mean profile transits smoothly from the porous layer to the fluid layer. The temporal evolution of several statistical quantities are plotted in figure~\ref{fig:conv_stats} and compared with the mean values given in~\citet{Reun21}. Both our simulations and those of~\citet{Reun21} generate the same statisitics.

\subsection{The 1D and 2D phase-changing problems}\label{subsec:phase}

In this section we simulate the 1D Stefan problem and the 2D flow of pure water with freezing and melting to demonstrate the applicability to simulating phase-changing flows.

\subsubsection{The 1D Stefan problem}

\begin{figure}
  \centering
  \includegraphics[width=0.9\textwidth]{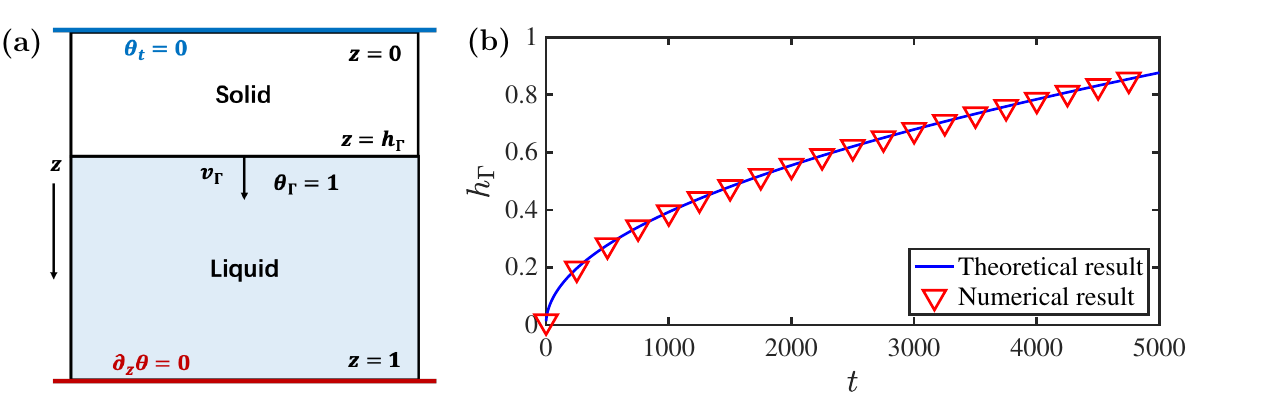}%
  \caption{The one-dimensional Stefan problem. (a) The schematic illustration of the problem setup. (b) Comparison of the temporal evolution of solid-liquid interface position $h_\Gamma(t)$ in simulation (symbols) with the analytical solution (blue line).}\label{fig:Ste1d}
\end{figure}

The configuration of 1D Stefan problem is depicted in figure~\ref{fig:Ste1d}(a). Initially the whole domain is occupied by liquid at freezing temperature $\theta_0=1$. Solidification is initiated by suddenly lowering the temperature of the top plate at $z=0$ to a sub-freezing value $\theta_t=0$. The liquid starts to freeze from $z=0$ and the solid-liquid interface propagates towards the other plate at $z=1$. The governing equation is pure diffusive in the solid region $0<z<h_\Gamma(t)$ and read
\begin{equation}
  \partial_t\theta(z,t) =\sqrt{\frac{Pr}{Ra}}\partial^2_z\theta(z,t) - St\partial_t\phi,  \qquad
  \partial_t\phi(z,t) =C_G\left(\theta-\theta_0\right).
\end{equation}
The boundary condition at interface $z=h_\Gamma$ is $\theta(h_\Gamma)=\theta_0$. The problem has the analytical solution with the interface position advances as 
\begin{equation}
  h_\Gamma(t) = 2\gamma\sqrt{\frac{Pr}{Ra\,t}}, \quad \mbox{with} \quad 
  \gamma\exp\left(\gamma^2\right){\rm erf}(\gamma) =\frac{1}{St\sqrt{\pi}}.
\end{equation}

We simulate the 1D Stefan problem by using our numerical method with a uniform grid of $N_z=240$. Other parameters are $St=1$ and $Pr/Ra=10^{-8}$. Figure~\ref{fig:Ste1d} (b) compares the temporal evolution of the interface position from our simulation with the analytical solution. The agreement demonstrates that the phase-field model accurately captures the dynamics of this classical solidification problem.

\subsubsection{The 2D freezing and melting of pure water with density anomaly}

In this section we simulate the freezing process in a layer of pure water. The computational domain is initially filled with water. The top plate is then kept constant at $T_t=-10^\circ\mathrm{C}$ which is below the freezing temperature of $T_0=0^\circ\mathrm{C}$. The bottom plate has the constant temperature of $T_b=10^\circ\mathrm{C}$ which is above the freezing temperature. The water near the top plate will freeze and form an ice layer at top, while convection flows appear near the hot bottom plate. The coupling between the freezing process and the convection flows together determine the final equilibrium state. Moreover, pure water under normal condition reaches its maximal density at the temperature $T_c=4^\circ\mathrm{C}$. Above and below $T_c$ the influence of temperature on density is opposite. The linear equation of state must be replaced by the following nonlinear one
\begin{equation}
  \rho = \rho_0\left(1-\alpha^*{\left\lvert T-T_c\right\rvert}^q\right),
\end{equation}
with $\rho_0=\rho(T_c)$ being the reference state and $q=1.895$.

The governing equations now read
\begin{subequations}
\begin{align}
    &\nabla\cdot\mathbf{u} = 0,	\\
    &\partial_{t}\mathbf{u}+\mathbf{u}\cdot\nabla\left(\frac{\mathbf{u}}{\phi}\right) 
    = -\nabla p + \sqrt{\frac{Pr}{Ra}}\left[\nabla^{2}\mathbf{u}-\frac{\phi}{Da\,f}\mathbf{u}\right]   
      +\phi{\left\lvert\theta-\theta_m\right\rvert }^{1.895}\mathbf{e}_z,  \\
    &\partial_{t}{\theta} = -\frac{\mathbf{u}\cdot\nabla{\theta}}{\alpha_c} 
      + \frac{\nabla\cdot\left(\alpha_k \nabla\theta\right)}{\alpha_c\sqrt{ Pr Ra }}
      - \frac{St}{\alpha_c}\partial_{t}\phi-\frac{\theta}{\alpha_c}\partial_t\alpha_c, 	\\
    &\partial_t\phi = -\mathbf{u}\cdot\nabla\phi +C_\phi\nabla^2\phi +C_G\left(\theta-\theta_0\right).
\end{align}
\end{subequations}
Note that the evolution of solid ice phase is tracked automatically by the phase field $\phi$. The Darcy number $Da=10^{-8}$ is sufficiently small so that the Darcy drag term is much stronger than the viscous stresses and the porous region with $\phi<1$ is effectively solid without any macroscale motions. In order to avoid the numerical instability caused by the extremely small porosity, a lower bound value is enforced as $\phi \ge \phi_{cr}=10^{-2}$. Our tests showed that this lower bound does not affect the simulation results once $\phi_{cr}$ is small enough. The two coefficients in the phase field equation are $C_\phi=10^{-8}$ and $C_G=10^{-1}$.

\begin{figure}
  \centering
  \includegraphics[width=\textwidth]{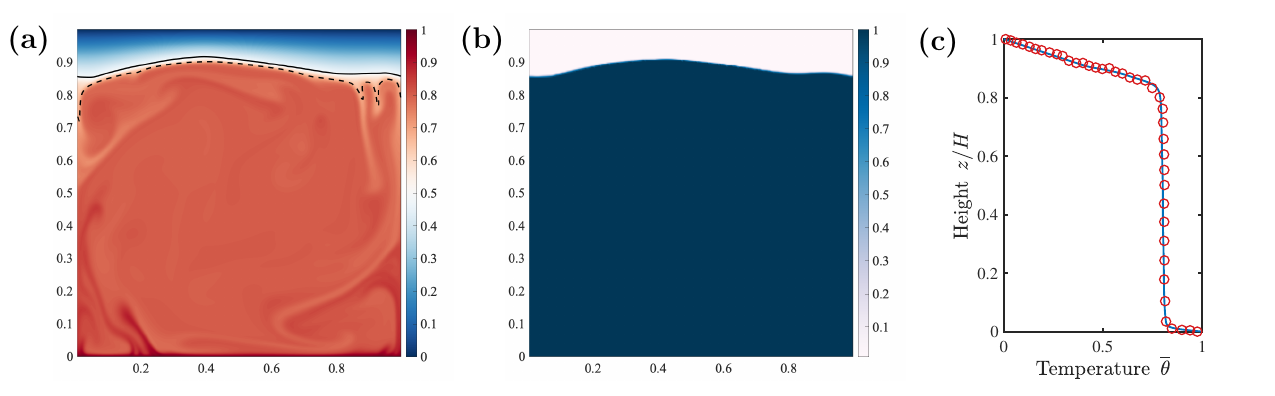}
  \caption{Validation against the two-dimensional freezing and melting experiment of pure water from \citet{Wang21}.\,(a) Snapshots of the temperature field $\theta$ from the present simulation at the final state.~(b) Corresponding snapshot of the porosity ($\phi$) field, where $\phi=1$ represents liquid water and $\phi=0$ represents solid ice.~(c) Vertical profile of the horizontally averaged temperature ${\left\langle\theta\right\rangle}_x$ at the final state. The results from the present simulation (line) are compared with the numerical data (symbols) from \citet{Wang21}.}\label{fig:waterice}
\end{figure}

We compare our results with the simulation results reported in~\citet{Wang21} where the Lattice Boltzmann Method (LBM) was employed. The control parameters are $Pr=10$ and $Ra=2\times10^9$. The simulation is run until the system reaches the final statistically steady state. Figure~\ref{fig:waterice}(a) shows a snapshot of temperature field, with the dashed line and solid line marking the maximal density contour and ice-water interface, respectively. The corresponding phase field distribution is given in figure~\ref{fig:waterice}(b). Our method captures the convection motion in the lower part of water region, the stably stratified layer between the maximal density line and ice front, and the heat conduction inside the ice region. For this case, the convection motions are strong enough to overcome the shielding effect of the stably stratified layer and cause the concave shape of ice-water interface. For the phase field, $\phi$ is nearly uniform and equal to the corresponding value of ice and water in the region away from the interface. Close to the interface, $phi$ smoothly transits from $\phi_{cr}$ in ice to $1$ in water in the direction normal to the interface.

Figure~\ref{fig:waterice}(c) compares the mean temperature profile obtained in our simulation with that given by the LBM simulation of~\cite{Wang21}. The agreement is satisfactory. We further compare the mean thickness $\overline{h}$ of ice layer and the Nusselt number $Nu$ which is the non-dimensional heat flux in the vertical direction. Our simulation gives $\overline{h}=0.106$ and $Nu=42.2$. While the results from~\citet{Wang21} are $\overline{h}=0.103$ and $Nu=43.2$.

\subsection{Growth of mushy ice during seawater freezing}\label{subsec:mushyice}

We now use the full solver to simulate the mushy ice layer development during the seawater freezing. The equation of state is more complex since both temperature and salinity should be considered and so does the density inversion effect. We take the same form as in~\citet{Du23}, namely
\begin{equation}\label{eq:eos_seawater_raw}
  \rho = b_1\left(1-b_2{\left\lvert T-T_m\right\rvert}^{1.895}\right).
\end{equation}
with $T_m=3.98(1-0.5266 S)$ being the temperature of maximum density. The other coefficients are defined as $b_1=\rho_0\,(1+b_0\,S)$ and $b_2=9.297\times 10^{-6}\,(1-0.02839\,S_i)$, with $\rho_0=999.972\,\mathrm{kg\,m^{-3}}$, $b_0=8.046\times 10^{-3}$, and $S_i$ being the initial salinity. Substituting the expressions for $b_1$ and $b_2$ into~\eqref{eq:eos_seawater_raw} yields a simplified form of the equation of state
\begin{equation}\label{eq:eos_seawater}
  \rho = \rho_0\left(1+b_0 S-b_2{\left\lvert T-T_m\right\rvert}^{1.895}\right).
\end{equation}
The governing equations are similar to~\eqref{eq:goveqn} with several necessary modifications. Due to the nonlinear equation of state, the characteristic velocity is now defined as $U=\sqrt{gb_2\Delta_T^{1.895}H\,}$ with $\Delta_T$ being the temperature difference across the domain height. The buoyancy force term in~\eqref{eq:db-ndim} changes to $\left(|\theta-\theta_m|^{1.895}-\Lambda s\right)\mathbf{e}_z$.

The mushy ice is a very unique porous medium which contains disconnected brine inclusions. Therefore, the phase field $\phi$ which is basically the volume fraction of liquid phase cannot be taken directly as the effective porosity. Instead, the effective porosity $\phi_e$, representing only the interconnected pore fraction, can be calculated according to the following model given by the percolation theory~\cite{Petrich06}
\begin{equation}
  \phi_e = \left\{
  \begin{aligned}
    & 0 &&\mathrm{for}\quad\phi\leq\phi_{cr}, \\
    & \alpha{\left(\phi-\phi_{cr}\right)}^\beta &&\mathrm{for}\quad\phi_{cr}<\phi\leq\phi_x, \\
    & \phi &&\mathrm{for}\quad\phi_x<\phi,
  \end{aligned}\right.
  \quad \text{with} \quad \phi_x = \frac{\phi_{cr}}{1-\beta}, \quad
  \alpha = \frac{1}{\beta}{\left(\frac{\beta\phi_{cr}}{1-\beta}\right)}^{1-\beta},
\end{equation}
where $\beta = 0.41$ and $\phi_{cr} = 0.054$ are sea-ice specific percolation parameters, and $\phi_x$ marks the transition to fully connected pores, respectively. The permeability follows the Kozeny–Carman relation $f(\phi_e) = \phi_e^3/{(1-\phi_e)}^2$ with reference permeability $K_0 = 3\times 10^{-11}\,\text{m}^2$. The porosity $\phi$ in \eqref{eq:db-ndim} is then replaced by the effective porosity $\phi_e$.

To capture the enhanced heat transfer through the mushy layer due to thermal dispersion, the effective thermal conductivity model~\eqref{eq:keff} with the dispersion coefficient given by
\begin{equation}
  \epsilon_{dis}(\phi) = \left\{
  \begin{aligned}
    &\epsilon_{d0} &&\text{for } 0 < \phi \leq \phi_{bn}, \\
    &\epsilon_{d0}{\left(\frac{1-\phi}{1-\phi_{bn}}\right)}^2 &&\text{for } \phi_{bn} < \phi \leq 1,
  \end{aligned}
  \right.
\end{equation}
where $\epsilon_{d0} = 10$ and $\phi_{bn}=0.9$. This formulation maintains dispersion in low-porosity regions while letting it decay smoothly to zero as $\phi\to 1$.

We simulate two cases reported in~\cite{Du23} with exactly the same settings. For Case I the initial salinity is $S_i=2.0\%$. The boundary temperature for the top plate is $T_t=-11.21^\circ\mathrm{C}$ and for the bottom plate is $T_b=4.79^\circ\mathrm{C}$. Case II has a higher initial salinity $S_i=3.5\%$. The boundary temperature is $T_t=-12.14^\circ\mathrm{C}$ and $T_b=3.86^\circ\mathrm{C}$, respectively. The two cases have the same temperature difference ratio $(T_t-T_0)/(T_0-T_b)$, or same superheat ratio. Take the average temperature of top and bottom plates as the reference temperature, the initial salinity $S_i$ as the reference salinity, respectively. The molecular diffusivity is chosen as $\kappa_s=7.16\times10^{-10}$~\cite{Gregg18}. Then the nondimensional parameters are $Pr=16.03$, $Ra=1.08\times10^8$, $St=4.44$, $\Lambda=17.58$, $Sc=2800$, and $Le=175$, respectively. The heat capacity ratio and thermal conductivity ratio between ice and water are $0.44$ and $4.29$.

\begin{figure}
  \centering
  \includegraphics[width=0.8\textwidth]{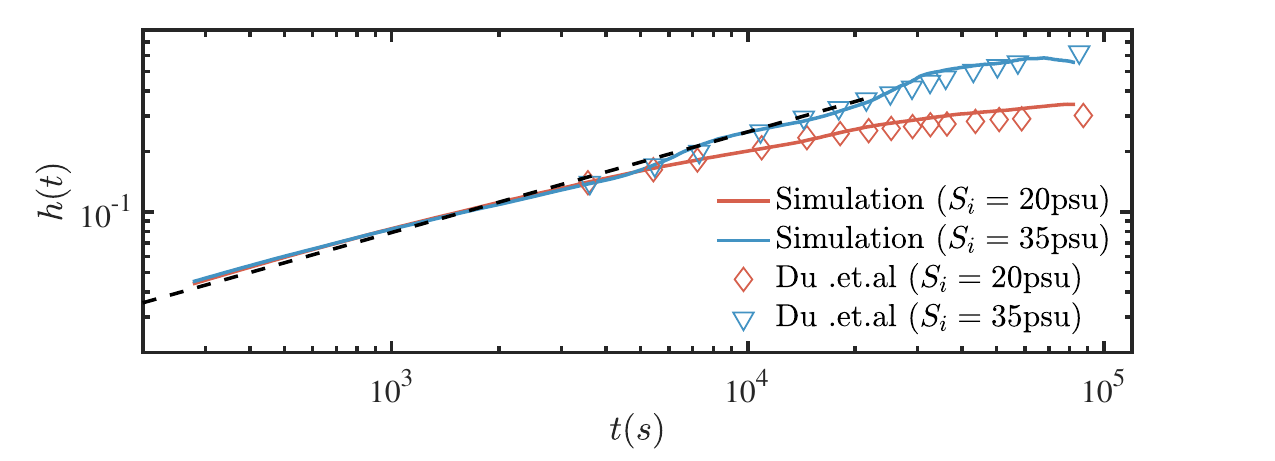}%
  \caption{Temporal evolution of mushy layer thickness $h_\Gamma(t)$ for initial salinity $S_i=2.0\%$ and $S_i=3.5\%$ under constant boundary temperature conditions. The numerical results from the present model (solid and dashed lines) are validated against the experimental data (symbols) from \citet{Du23}.}\label{fig:mushy_mt}
\end{figure}

\begin{figure}
  \centering
  \includegraphics[width=\textwidth]{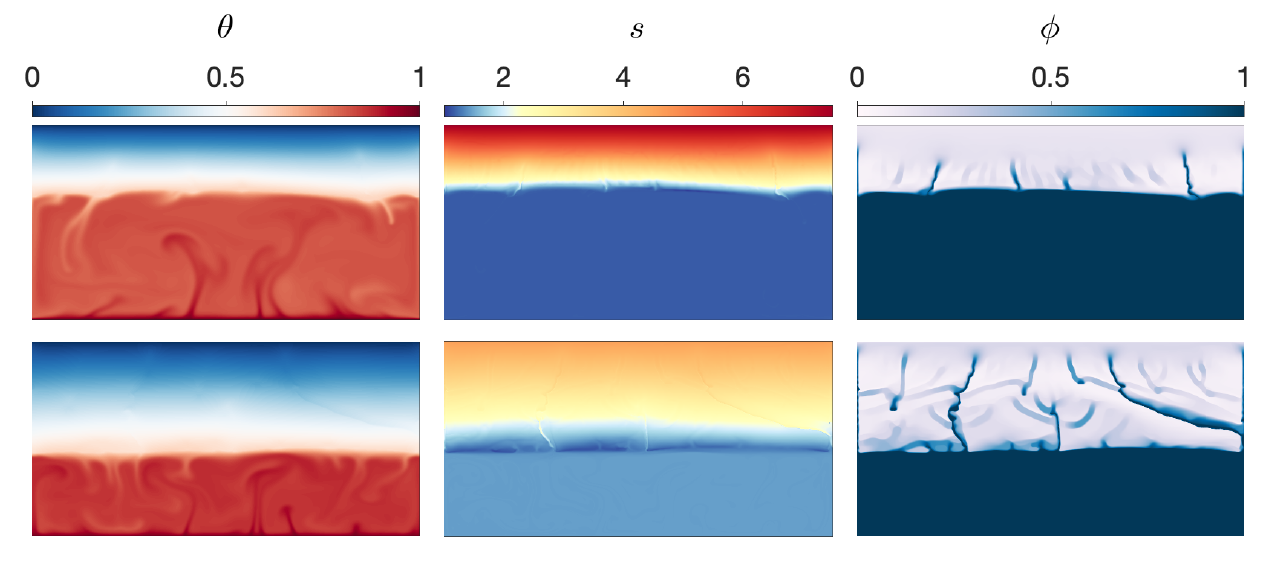}%
  \caption{The scalar fields depict the mushy ice layer and the fluid convection of Cases I (top row) and II (bottom row) at $t=83000$ seconds. From left to right: temperature $\theta$, salinity $s$, and porosity field $\phi$, respectively.}\label{fig:mushy_field}
\end{figure}

We first compare the temporal evolution of the mushy ice layer thickness $h$ with the experimental results in figure~\ref{fig:mushy_mt} for two cases with different initial salinity. Our numerical method is capable of capturing the temporal evolution of $h$ for a very long time period for both cases. In figure~\ref{fig:mushy_field} we present the flow fields of the two cases at $t=83000$ seconds, by showing the contours of temperature, salinity and phase field $\phi$. The whole evolution history for the flow fields of two Cases can be seen in the supplementary movies. Evidently, the mushy ice layer is very inhomogeneous and exhibits very dynamic evolution. Especially, thin channels form and drain salinity from the ice layer to the liquid layer below. The salinity plumes originated from exits of these channels at interface enhance the local convection motions. Also, for Case II with higher initial salinity, the channels are more pronounced. All these findings are consistent with the experimental observations. 

Although our numerical method generates satisfactory results for the two cases of seawater freezing, there are quite some parameters which should be fixed either by theory or by calibration according experimental measurements. Especially, measuring the micro structures and macro properties of mushy ice is extremely challenging. As the end of the subsection, we briefly discuss the influence of several vital parameters in our numerical simulations.

The value $K_0=3\times 10^{-11}$ used here is smaller than that suggested by \citet{Du23}. This is because the permeability reported by \citet{Du23} was derived based on the average porosity of the final ice layer. In reality, however, the heterogeneous pore structure of sea ice which characterized by local high-porosity brine channels, is significantly enhances the effective permeability. In simulations, if $K_0$ is set too high, brine is expelled too rapidly in the initial stage, leading to premature porosity reduction and suppressed convection, slowing ice growth below the observed rate. Conversely, if $K_0$ is too low, brine rejection is suppressed throughout, also yielding slower growth. 

We set $\epsilon_{d0}=10$ which is significantly larger than values derived from solid-sphere theory for packed beds. This enhancement is physically motivated by the dendritic microstructure that develops during sea-ice formation: a large number of brine channels and dendritic ice plates align with the temperature gradient, creating highly conductive pathways that markedly accelerate heat transfer. In simulations, if $\epsilon_{d0}$ is set too low (e.g., close to the classical packed-bed value), the effective thermal conductivity within the mushy layer becomes insufficient, leading to much slower ice growth rates that fall far below the experimentally observed evolution.

The phase-field diffusion coefficient $C_\phi$ also requires careful selection to balance numerical stability and physical fidelity. Basically, $C_\phi$ controls the small scales of phase field. If $C_\phi$ is too large, the porosity field becomes smooth and the fine brine-channel structures that are essential for salt and heat transport within sea ice disappear. If $C_\phi$ is set too small, large local gradients develop in phase field which alters the effective permeability $\mathcal{K}(\phi)$ through the highly nonlinear Kozeny–Carman relation. In the present simulations we set $C_\phi=2\times 10^{-6}$, which generates the characteristic scale of brine channels observed in the experiments.

The phase-change kinetics coefficient $C_G$, which governs the response of porosity to thermal under-cooling, must be chosen to reproduce the experimental freezing rate while maintaining numerical stability. If $C_G$ is too small, the phase transition proceeds too slowly, and the diffusion term $C_\phi\nabla^2\phi$ dominates and smears out the brine-channel microstructure. failing to capture the characteristic sea-ice morphology observed experimentally. On the other hand, if $C_G$ is too large, freezing becomes unrealistically rapid. This leads to the formation of grid-scale features in the porosity field which distort the effective permeability and heat transport. In our simulations, we use $C_G = 10^{-1}$ which yields ice-growth timescales consistent with the experiments.

\section{Conclusions}\label{sec:con}

In summary, we present a numerical framework which can be used to simulate incompressible flows in coupled fluid-porous-medium systems with multiple scalar components and phase-changing process. Not only the flow motions within fluid and porous medium can be modeled with the current method, but also the dynamic evolution of the porous medium itself can be simulated. Due to the fact that the method can treat porous medium with very different porosity, solid phase can also be effectively modeled by a porous medium with very low porosity. Therefore, the method developed here is very versatile and suitable for various flows.

A complete set of governing equations are constructed to serve as the base of numerical methods. The macroscopic motions of incompressible flows in fluid and/or porous medium are described a unified form of Darcy-Brinkman equation which, with the assistant of a phase field, can automatically accommodate different types of medium. The temperature and solute equations are the standard advection-diffusion type with spatiotemporally varying diffusivity and source terms associated to phase-changing process. The phase field follows also a advection-diffusion equation and its dynamics is driven by the phase-changing a binary system. 

A numerical method is then proposed to solve the dynamics system for velocity, phase field, temperature and solute concentration. The method employs the fractional time-step scheme and the second-order finite difference discretization. A third-order Runge-Kutta scheme is used for temporal integration. The nonlinear and source terms are treated by the explicit Adams-Bashforth type of scheme and the diffusion terms by the semi-implicit Crank-Nicholson type of scheme, respectively. The diffusion terms with non-uniform diffusivity are split into the part with uniform diffusivity equal to the maximal value over the field, and the residual part taking care of spatial variation of diffusivity. The former is treated by the standard semi-implicit scheme, while the latter by the explicit scheme, respectively.

The numerical method is then applied to a series of model problems and validated against the existing simulations and experiments. For the coupled fluid-porous-medium system, the turbulent flows over the porous substrate and the convection flows in the two-layer system are simulated. The phase-changing module is tested by the 1D Stefan problem and 2D flows of freezing in pure water. The results of these simulations are compared with the existing simulations which used different numerical methods or the analytical solution when available. Finally, the growth of mushy ice in seawater is simulated for two different cases and the results are compared with experiments at exactly the same conditions. For all comparisons satisfactory agreement is obtained.

Given the versatility of the current method, improvements are needed in several aspects. Systematic validations are currently underway for simulating turbulent flows above porous substrate and the mushy ice growth over much widely parameter range. The coefficients in the numerical method and especially those in the mushy layer growth should be carefully calibrated by combining theoretical analyses, new experimental measurement, and pore-size resolved simulations. The last one is the most promising one as such numerical method has been available, such as those developed by~\citet{Wei25}. By conducting pore-size resolved simulation, a more reliable macroscale constitutive model for the mushy layer can be established, which will greatly improve the current method. 

\section*{Acknowledgments}

This work is supported by the NSFC Young Scientists Fund (A) under grant No.~12525209 and the NSFC Excellence Research Group Program for `Multiscale Problems in Nonlinear Mechanics' under grant No.~12588201.

\bibliography{PorousBoundaryMethod}

\end{document}